\documentclass{aa}  
\usepackage{graphicx}
\begin{document}

   \title{Updated pulsation models for BL Herculis stars}

   \author{Marconi, M.
          \inst{1} 
          \& Di Criscienzo, M.\inst{1}\inst{2}
          }

   \offprints{Marconi, M.}

   \institute{INAF-Osservatorio Astronomico di Capodimonte,\\
              via Moiariello, 16,I-0080131, Naples, Italy
         \and
             Universit\'a di Roma ``Tor Vergata''\\
             via della Ricerca Scientifica, 1, Rome, Italy
}
   \date{Received ????; accepted ???}
\authorrunning{Marconi \& Di Criscienzo}
\titlerunning{Updated pulsation models for BL Herculis stars}
\abstract{{\it Context} Population II pulsating
variables play a
     relevant role both as distance indicators and as tracers of the
     properties of old stellar populations.\\
     {\it Aims} In this paper we present an updated and homogeneous
pulsational scenario
for a wide range of stellar parameters typical of BL Her stars i.e.,
Population II Cepheids with periods shorter than 8 days.
    {\it Methods} To this purpose, we adopt a nonlinear convective
 hydrodinamical code
  evaluate the  stability and full amplitude behaviour of an extensive set of BL Her pulsation
models.
Various assumptions of mass, luminosity and metallicity
consistent with the most recent evolutionary prescriptions, are adopted.
  {\it Results}
We obtain the theoretical instability strip for both fundamental and first overtone pulsators and
present a detailed atlas of
light/radial-velocity curves. Some relations for the boundaries of
the instability strip and
for the dependence of the absolute magnitude on period, mass and color, as well as the fundamental
period-amplitude relations are
derived.Finally, we  provide the theoretical period-radius relation for BL
Her and find
that it is in
excellent agreement with the empirical relation by Burki \& Meylan and
consistent with the one holding for shorter periods for
RR Lyrae stars.
                                                                                                   
}
\maketitle
%
%________________________________________________________________

\section{Introduction}
The BL Her variables form a small but interesting group of radially pulsating 
Population II stars with period $P$ in the range 0.8-3 days. 
They are observed in globular clusters with few RR Lyrae stars
and blue horizontal branch (HB) morphology and they are 
brighter than RR Lyrae but fainter than Anomalous Cepheids, for a 
fixed period. They are currently interpreted as 
central He-burning low mass
($M\le 0.6M_{\odot}$) stars which, after their main core He-burning phase 
spent in the blue side of the Zero Age Horizontal Branch 
(ZAHB), evolve toward the Asymptotic Giant Branch (AGB) 
crossing the instability strip at 
luminosities higher than the  RR Lyrae level 
(see Gingold 1985; Bono et al. 1997a, Caputo
1998).\\
The properties of these objects have recently been reviewed by Wallerstein (2002) 
and Sandage \& Tammann (2006). 
From the theoretical point of view, Buchler \& Moskalik (1992) 
and Buchler \& Buchler (1994) 
presented a pulsational study of models with $P\le$ 3 days 
 based on a linear and nonlinear radiative analysis. 
These authors found evidence for a resonance between the 
fundamental mode and the second overtone, similar to the one often 
invoked as the explanation of the {\it Hertzsprung progression} phenomenon in 
Classical Cepheids (see Bono, Marconi, Stellingwerf 2000 for a detailed discussion).
Buchler \& Buchler (1994) also evaluated the pulsation properties of first overtone (FO)
models and found that the first overtone blue edge (FOBE) was 
very close ($\le$ 100 K) 
to the fundamental (F) one, producing a very narrow regime of FO-only pulsation. 
Moreover, FO models were found to show very low pulsation 
amplitudes but it seemed to be hard to discriminate the pulsation mode on the basis 
of their Fourier coefficients.
By relying on a  nonlinear convective pulsation analysis, Bono et al. (1997a) found  
a good agreement between the predicted and the 
observed BL Her instability strip boundaries  and suggested 
that these pulsators are pulsating in the F mode with a
typical mass of 0.52-0.59 $M_{\odot}$. Moreover these authors predicted a 
Period-Luminosity-Amplitude (PLA) relation in the B band.\\
However the Bono et al. (1997a) nonlinear convective approach  
was limited to a quite restricted range 
of stellar parameters and adopted an old input physics 
both in the evolutionary and the pulsational computations
(see Bono \& Stellingwerf 1994 and Bono et al. 1997  for details).
In this paper we present an updated and homogeneous pulsational scenario 
for a wide range of stellar parameters typical of BL Her stars.
In an accompanying  paper (Di Criscienzo et al. 2006) we will discuss 
the connection  between the pulsational scenario and the predictions 
of updated evolutionary 
models in order to compare theory
 with observations. 
 
The organization of the present paper is the following: in Sect.2 we present 
the new computed models. In Sect. 3 we
 deal with  predicted light/velocity curves and visual amplitudes.
Same important relations are given in Section 4, whereas the 
conclusions close the paper.
%__________________________________________________________________

\section{The stellar pulsation models }
\begin{table}
\caption{Input parameters of the computed BL Her models. An 
helium abundance Y=0.24 has been adopted.}                
% title of Table
\label{table:1}      % is used to refer this table in the text
\centering                          % used for centering table
\begin{tabular}{c c c c c c c}        % centered columns (4 columns)
\hline\hline                 % inserts double horizontal lines
Z& M/M$_{\odot}$ & LogL/L$_{\odot}$& FOBE&FBE&FORE&FRE \\    % table heading 
\hline                        % inserts single horizontal line
0.0001& 0.60 & 1.95 &-&6850&-&5750\\
      &      & 2.05 &-&6750&-&5650\\
      &      & 2.15 &-&6750&-&5550\\     % inserting body of the table
      & 0.65 & 1.91 &6950&6850&6100&5750\\
      &      & 2.01 &6750&6850&6300&5750\\
      &      & 2.11 &-&6750&-&5550\\
\hline
0.001 & 0.50 & 2.11 &-&6650&-&5450\\
      &      & 2.41 &-&6350&-&5150\\
      & 0.55 & 1.81 &6875&6850&6400&5650\\
      &      & 1.91 &-&6850&-&5550\\
      &      & 2.01 &-&6750&-&5450\\
      & 0.65 & 1.81 &7050&6750&6650&5750\\
      &      & 1.91 &6850&6750&6150&5650\\
      &      & 2.01 &6650&6850&6350&5650\\
\hline
0.004 & 0.55 & 1.81 &-&6950&-&5750 \\
      &      & 1.91 &-&6850&-&5650 \\
      &      & 2.01 &-&6750&-&5450 \\
\hline                                   %inserts single line
\end{tabular}
\end{table}
\begin{figure*}
      \centering
      \includegraphics[width=13cm]{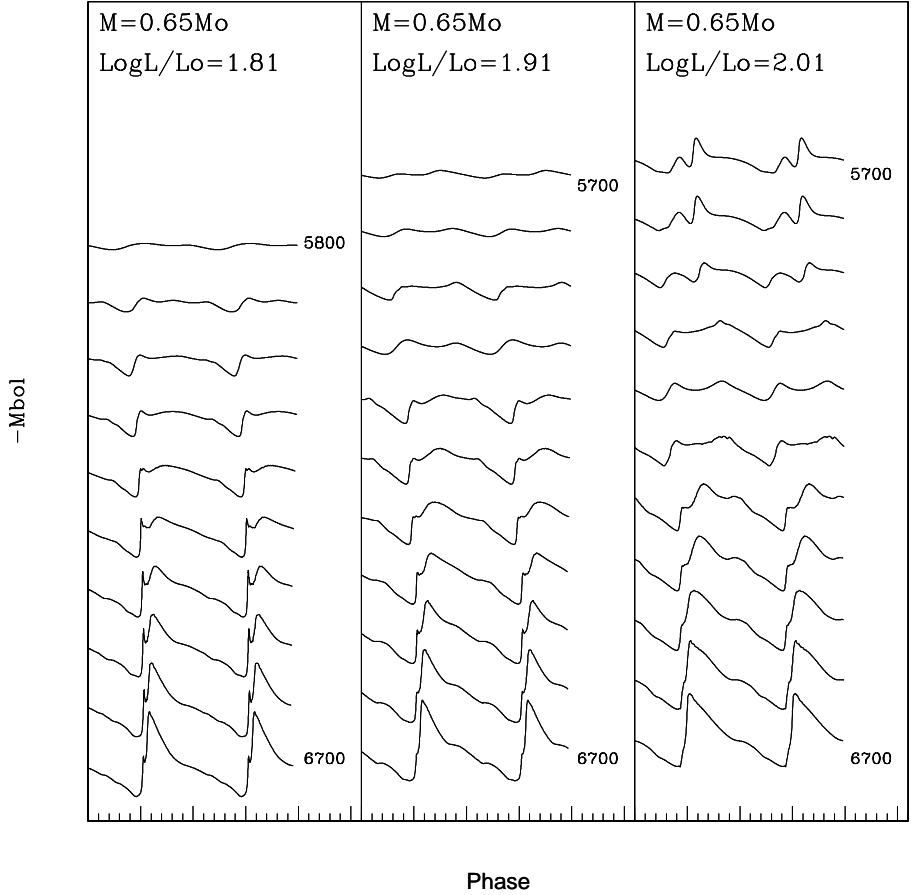}
      \caption{Theoretical bolometric light curves
               for a subsample (Z=0.001 and M=0.65 M$_\odot$) of fundamental models.The luminosity levels are labeled.
              }       \label{curve}
   \end{figure*}
\begin{figure*} 
      \centering
      \includegraphics[width=13cm]{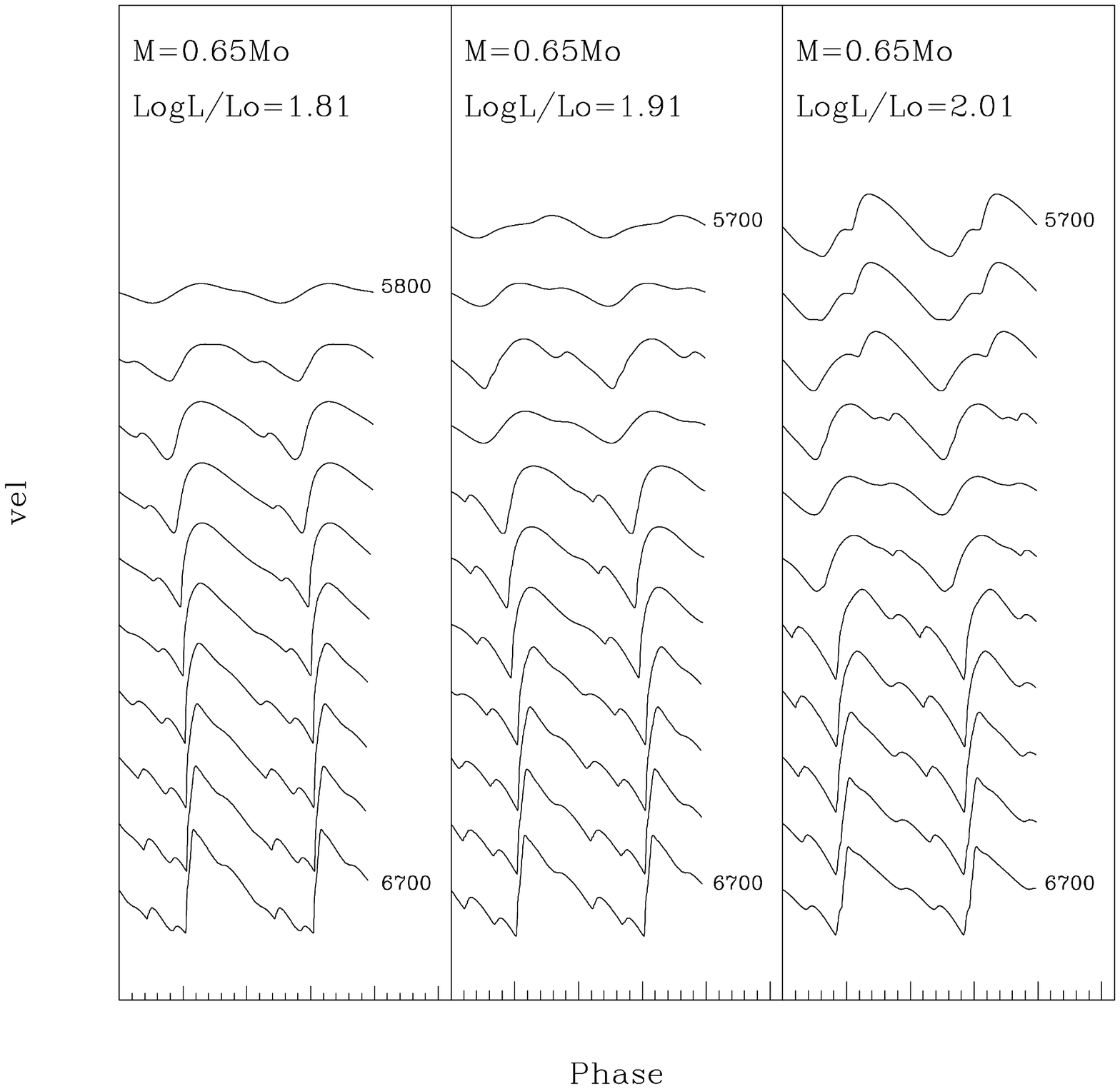}
      \caption{The same as Fig. 2 but for theoretical radial velocity curves.
              }
         \label{vel}
   \end{figure*}
To study BL Her stars, we have adopted the same pulsation code and the same 
physical and numerical assumptions used for RR Lyrae  
(Bono $\&$ Stellingwerf 1994, Bono, Castellani \& Marconi 2000, Marconi et al.2003, 
Di Criscienzo et al. 2004) and Anomalous Cepheid models (Marconi et al. 2006). 
Model sequences have been computed as one parameter families with constant 
luminosity, mass and chemical composition, by varying the effective temperature 
with a step of  100 K. The various assumptions of mass, luminosity and metallicity 
are consistent with the most recent evolutionary prescriptions 
(see Pietrinferni et al. 2004, 2006) and are reported in the first three columns 
of  Table 1.

Linear regression through the model's parameters allows us to derive analytical relations connecting the period of models to the intrinsic stellar parameters, namely mass, luminosity and effective temperature, i.e:\\
{\small{\begin{eqnarray}
\nonumber logP&=&11.579 (\pm 0.015) + 0.893 \log \frac{L}{L_{\odot}}-0.89  \log \frac{M}{M_{\odot}}\\
&-& 3.54 \log T_{\mathrm{eff}}
\end{eqnarray}}
for fundamental pulsators and:
{\small{\begin{eqnarray}
\nonumber logP&=&10.784 (\pm 0.003) + 0.806 \log \frac{L}{L_{\odot}}-0.66  \log \frac{M}{M_{\odot}}\\
&-& 3.31 \log T_{\mathrm{eff}}
\end{eqnarray}}
for FO ones.

It is well known that radial pulsation occurs in a quite well-defined region of 
the HR diagram, the so called instability strip, which depends on 
the intrinsic parameters of pulsators. For each given metallicity, mass and luminosity, 
we have calculated the  maximum and minimum effective temperature 
for the onset of either fundamental or first-overtone pulsation, 
as reported in the last four columns of Table 1, where FOBE (FORE) and FBE (FRE) are the blue (red) limit for  first overtone and fundamental pulsator respectively.
It is of importance to note that  our calculations 
confirm the earlier suggestion by Bono et al. 1997(a,b,c) that, for a given 
helium content and  
mass, there exists a "intersection" luminosity $L_{IP}$ where FOBE=FBE and that 
above this luminosity only the fundamental mode has a stable nonlinear limit cycle.   
As shown by Di Criscienzo et al. 2006, the evolutionary models 
show that  BL Her variables occur at luminosity higher 
than  this intersection point: therefore, in the following we will 
consider only fundamental models. As for 
the red boundary of the fundamental mode, it is worth mentioning 
that it is expected to depend on the efficiency of convection 
and, in particular, on the value of the mixing length parameter 
$l/H_{p}$ we assume  to close the nonlinear system of dynamical and convective 
equations (see Stellingwerf 1982 and Bono \& Stellingwerf 1992, 1994 
for details on the treatment of convection in the adopted hydrodynamical code). 
To test this occurrence, we have computed additional models with 
$l/H_{p}$ increased from 1.5 to 2.0 and we confirm  
the general trend shown by  
RR Lyrae (Di Criscienzo et al. 2004) and Classical Cepheid models  
(Fiorentino et al. 2006, in preparation) in that  
the FRE moves toward higher effective temperatures 
as $l/H_{p}$ increases. In addition, we find that this effect is 
less important with increasing the luminosity. In particular, 
for $M=0.50M_{\odot}$ and $Z$=0.001 the FRE blueshift 
is of about 200 K at $logL/L_{\odot}=2.01$  but reduces to less than 
100 K for  $logL/L_{\odot}=2.41$. 
This occurrence is due to the reduced density  values, and 
in turn to the increase of the required superadiabaticity,  
in the outer layers of stellar structures with low mass and very high luminosity.
%_________________________________________________________
\section{Light curves and pulsation amplitudes}
For each investigated model, the non linear analysis 
provides the variation of relevant parameters, namely luminosity, 
radius, radial velocity, effective temperature and gravity, along a pulsational cycle.
A subsample of bolometric light curves ($Z$=0.001 and $M=0.65M_{\odot}$ 
and the labeled luminosity levels) is shown in Figure~\ref{curve}, 
while Figure~\ref{vel} exhibits the corresponding radial velocity curves 
as a function of pulsation phase. 
All these curves show a large variety of shapes, 
which is perhaps the most striking feature of BL Her models 
as already suggested by Moskalik and Buchler (1994). 
In all the sequences (in particular for radial velocity curves), 
the presence and the progression of the main  bump is evident in analogy 
of what observed for  Classical Cepheids with periods around 10 days 
(the well known Hertzsprung progression).
A detailed investigation of this issue, through the study of Fourier parameters,  
will be addressed in a future paper.

In order to compare theoretical results with observations, 
the bolometric lights curves have been transformed into the photometric 
bands $UBVRIJK$, using bolometric corrections and temperature-color relations 
provided by Castelli, Gratton $\&$ Kurucz (1997a,b). 
In this way, light-curve amplitudes and mean absolute magnitudes, 
either intensity-weighted  $\langle M\rangle$ or magnitude-weighted $(M)$,  
are derived in various photometric bands. 
After having tested that intensity-weighted magnitudes approximate better the static 
values than magnitude-weighted ones, 
similarly to what found for RR Lyrae (see e.g. Marconi et al. 2003) 
and Classical and Anomalous 
Cepheids (Caputo, Marconi, Ripepi 1999, Marconi et al. 2004), 
in the following we give the predicted relations with 
$\langle$ M$\rangle$ also because empirical 
investigations usually provide this type of mean values\footnote{Similar relations involving magnitude-weighted mean values are available 
upon request to the authors}.

\begin{figure*}
      \centering
      \includegraphics[width=13cm]{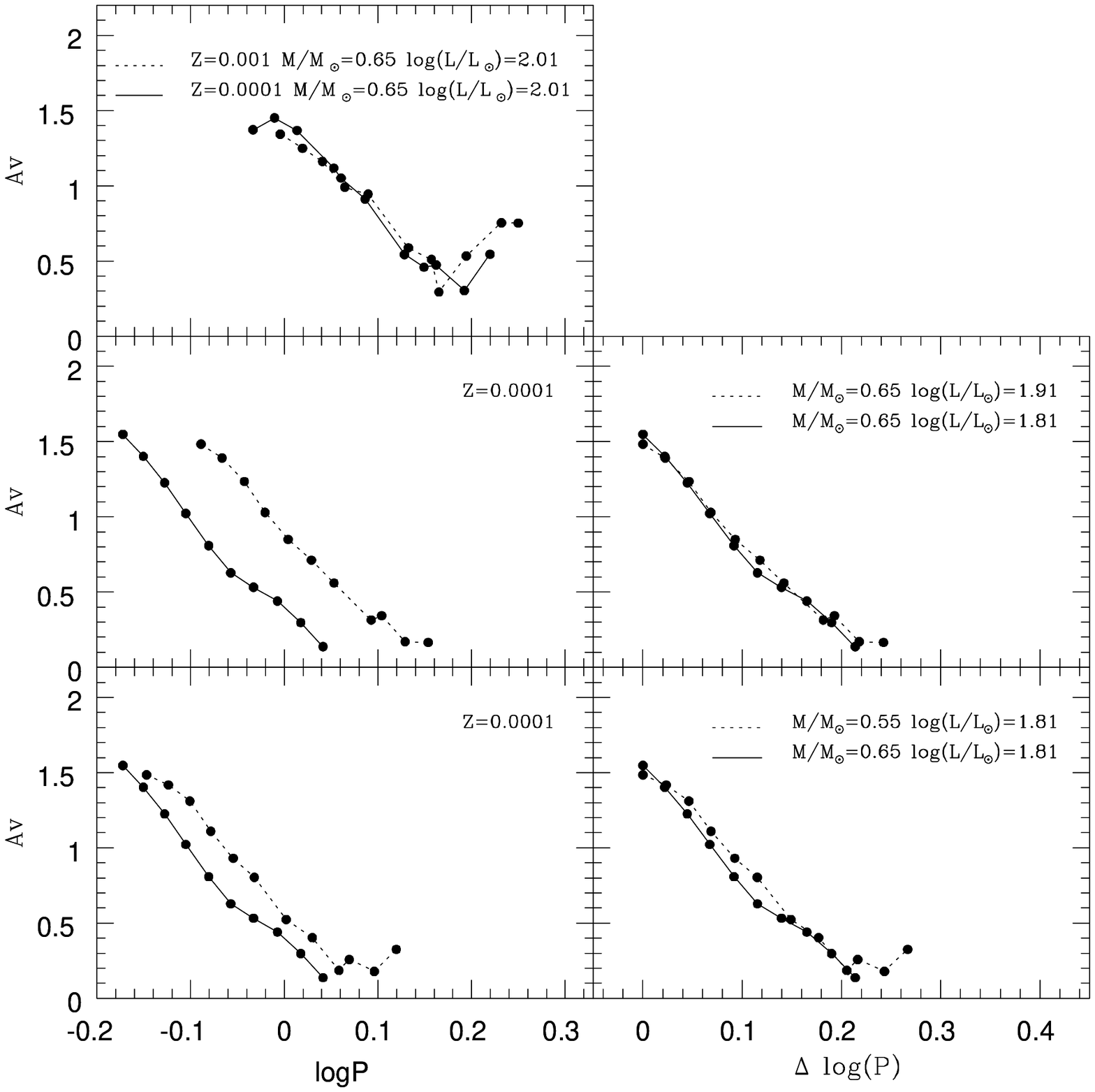}
      \caption{(Right panels)Visual amplitudes versus periods for selected models  varying the metallicity at fixed mass and luminosity (upper panel), 
   varying the luminosity at fixed mass and metallicity (Z=0.001, intermediate panel) and varying the mass at fixed luminosity and metallicity (Z=0.001, lower panel).(Left panels) The same but visual amplitudes are now plotted versus ${\Delta}$ log(P)=log (P)- log(P$_{FOBE}$)
              }
         \label{amp}
   \end{figure*}
In  Figure~\ref{amp}, we show the behaviour of visual amplitudes 
as a function of the period for different  values of metallicity, mass and luminosity. 
As for RR Lyrae models, we find that for fixed period the amplitude increases 
as the luminosity increases  and as the mass decreases (middle and bottom left panels, 
respectively), while it remains quite constant in the considered interval 
of metallicity, at fixed mass and luminosity (top left panel). 
We notice that part of the observed behaviour in the period-amplitude diagram is due to the dependence of period on mass and luminosity, as simply derived even from a simple linear adiabatic approach. The change in the pulsation in  amplitude is mainly related to the distance from the FBE, as shown in the two right panels of the same figure.
The deviation from linearity of the highest luminosity level 
models (top panel of Figure~\ref{amp}) is related to the complex coupling between pulsation and convection for these low density and cool structures.
(see also Bono et al. 1997c for a similar behaviour in high luminosity RR Lyrae stars).
However in the range of linearity we obtain:
{\small{\begin{eqnarray}
\nonumber \log P&=&-0.033(\pm 0.034)-1.15\log\frac{M}{M_{\odot}}-0.475\log <M_V>\\
&-&0.195 A_V
\end{eqnarray}}
We notice that this dependence of the period on amplitude is, within the errors, in agreement with those found in Di Criscienzo et al. 2004.
%_________________________________________________________
\section{Same relevant relations}
\begin{figure}
      \resizebox{\hsize}{!}{\includegraphics{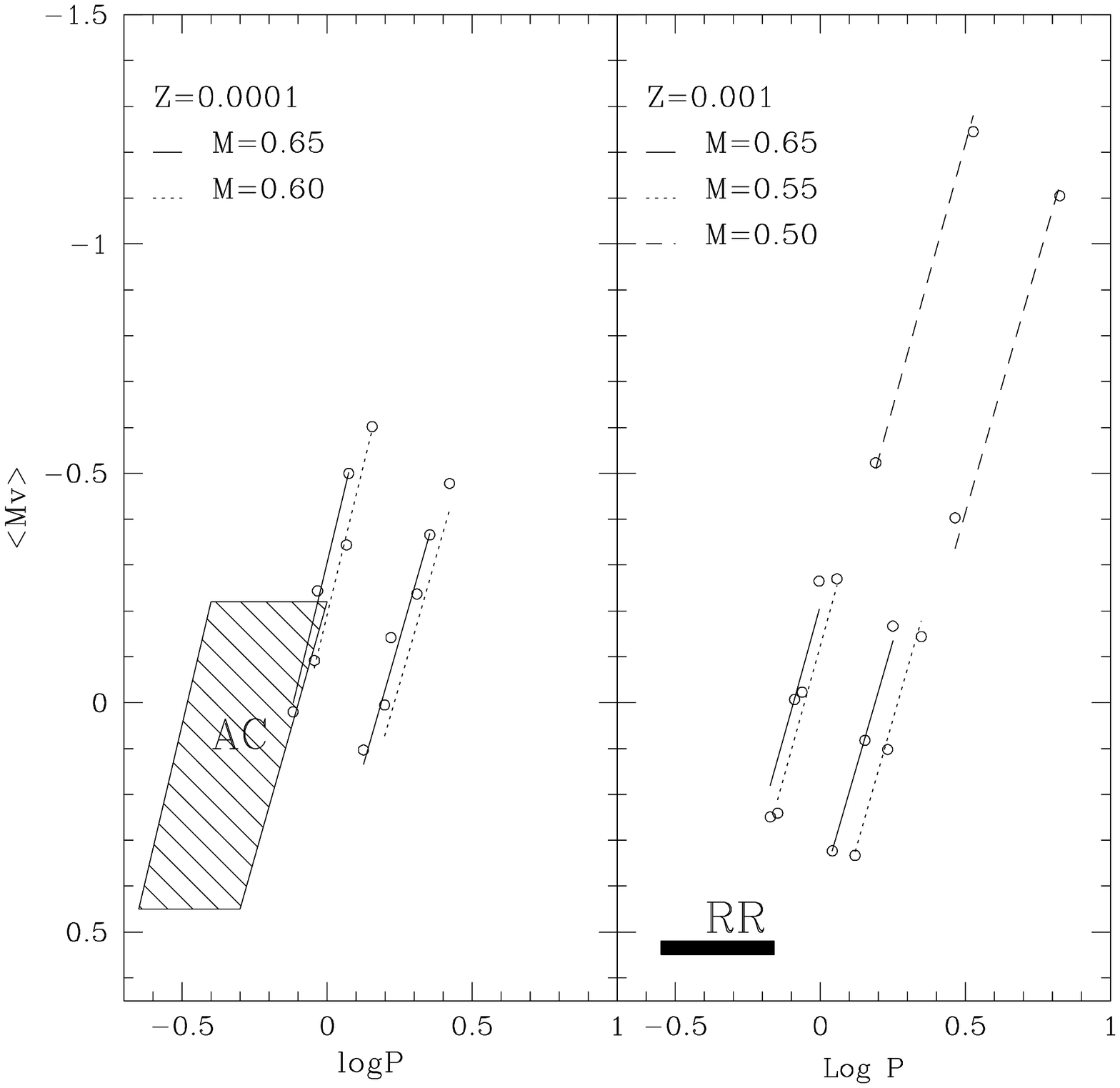}}
      \caption{Dependence of the boundaries of the pulsation region on luminosity in the $M_V$-$\log$P plane for the labeled values of mass and metallicity. In the left
      panel we also report the AC instability strip for a mass of 1.3M$_{\odot}$ (Marconi et al. 2004) while in the right one we show the predicted
      RR Lyrae instability strip for Z=0.001 as obtained by synthetic simulations.
}             
         \label{bordi}
   \end{figure}
\begin{figure}
      \resizebox{\hsize}{!}{\includegraphics{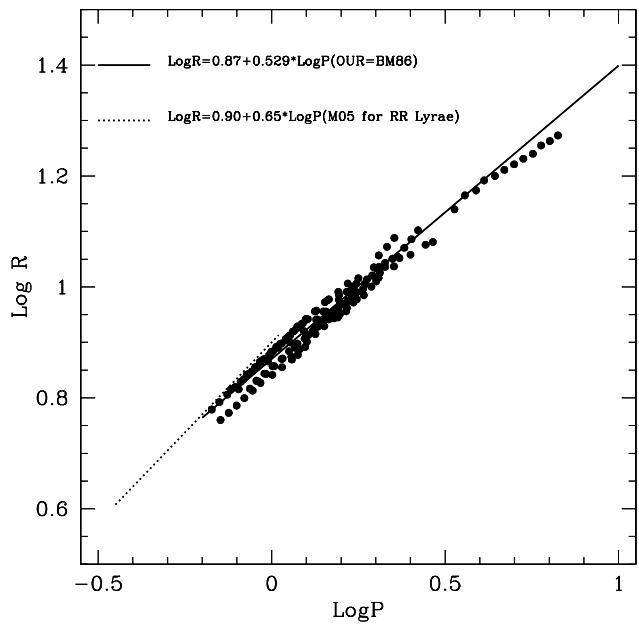}}
      \caption{Period-Radius relation for our models. The solid line is the linear regression throught the entire set of BL Her models (filled circle)  which is exactly the same found by Burki \& Meylan, 1996 (BM96), while the dashed one  is the PR relation  obtained by Marconi et al. 2005 (M05) for RR Lyrae stars.
              }
         \label{pr}
   \end{figure}

The linear regression through the data reported in Table 1 
provides the following analytical relations:
{\small{\begin{eqnarray}
\log{T_e}(FBE)=3.912(\pm 0.005)-0.035 \log \frac{L}{L_{\odot}} +0.048\log\frac{M}{M_{\odot}}\\ \log{T_e}(FRE)=3.925(\pm 0.005)-0.075 \log \frac{L}{L_{\odot}} +0.118\log\frac{M}{M_{\odot}}
\end{eqnarray}}
with a rms of 0.005 , valid on the adopted range of metallicity and stellar mass.
The location of these edges in the $M_V$-log$P$ plane is 
presented in Figure~\ref{bordi} and the corresponding analytical relations are:

{\small{\begin{eqnarray}
\nonumber <M_V(FBE)>&=&-0.36(\pm 0.04 )-1.5 5\log\frac{M}{M_{\odot}}-2.37\log P\\
&+&0.05(\pm0.03) \log Z\\
\nonumber <M_V(FRE)>&=&0.14(\pm 0.03)-2.25\log\frac{M}{M_{\odot}}-2.17 \log P\\
&+&0.05 (\pm0.02) \log Z
\end{eqnarray}}

From the same Figure, we note that the BL Her instability strip is  
the extension at higher luminosities of the
 fundamental instability region for RR Lyrae stars, whereas 
it does not overlap the one for Anomalous Cepheids
(ACs) since the AC FRE is hotter than the BL Her FBE (see also  
Caputo et al. 2004). This result, also related to the significant mass difference between the two classes of pulsators, is fully consistent with the empirical evidence 
that ACs are brighter than BL Her stars, for a given period.

The natural outcome of the period relation (Eq. 1) in the observational plane 
is the Period-Magnitude-Color relation
in which the pulsation period is correlated with the pulsator 
absolute magnitude and color, for a given mass.
As a matter of example, linear interpolations through the results give:
{\small{\begin{eqnarray}
\nonumber 
\langle M_V\rangle&=&-1.22(\pm 0.05)-2.60\log P-3.03[\langle M_B\rangle-\langle M_V\rangle]\\ 
&-&1.10 \log \frac{M}{M_{\odot}}\\
%\end{eqnarray}
%\begin{eqnarray}
\nonumber \langle M_V\rangle&=&-2.08(\pm 0.02)-2.88\log P-4.13[\langle M_V\rangle-\langle M_K\rangle]\\ 
&-&2.14 \log \frac{M}{M_{\odot}}
\end{eqnarray}}
with no dependence on metallicity. According to these relations, 
for a sample of variables at the same distance and with the same reddening,
e.g., for variables in a given globular cluster or galaxy with no differential reddening,
one could  estimate the mass range spanned by the variables from 
the observed luminosity range,
once periods and intrinsic colors are firmly known. However, these relations 
have the disadvantage 
of depending on reddening uncertainties and for this reason, as discussed in several 
earlier papers (e.g., Madore 1982, Di Criscienzo et al. 2004, Udalski et al.1999),   
Period-Wesenheit relations are widely used, with the Wesenheit functions being 
by definition reddening insensitive. The full set of 
the predicted Period-Wesenheit and Period-Magnitude relations,  
including the 
evolutionary properties of the BL Her stars, 
are reported in Di Criscienzo et al. (2006).
Figure~\ref{pr}  shows the behavior of the predicted radii for fundamental 
pulsators, obtained by averaging the
theoretical radius curve. 
The solid line is the linear regression through the entire set of models 
($PR$ relation):
\begin{equation}
\log R = 0.87 (\pm0.01)+ 0.529 (\pm0.006) \log P
\end{equation}
where $R$ is in solar units.
This relation is in agreement with the empirical 
$PR$ relation obtained by Burki \& Meylan, 1986.

In the same Figure we have also reported the theoretical 
$PR$ relation recently obtained by Marconi et al. (2005) for RR Lyrae. 
This very good agreement
between the theoretical PR relations obtained for RR Lyrae and BL Her stars 
supports earlier suggestions by Burki \& Meylan (1986) concerning 
the similarity of the $PR$ relations for these two classes of Pop. II variables, 
and is consistent with 
the results by Caputo et al. (2004) concerning the behaviour of RR Lyrae and BL Her 
in the Period-Magnitude diagram. 
\section {Conclusion and final remarks}
This work is part of a larger project that has the scope to study, from the 
theoretical point of view, Population II variables in Globular Clusters and
dwarf galaxies.
In the last few years we have provided the results for RR Lyrae stars (Marconi et
al
2003, Di Criscienzo et al.2004) and Anomalous Cepheids (Marconi,
Fiorentino \& Caputo,2004, Caputo et al. 2004, Fiorentino et al. 2006). 
The present paper, together with the investigation by Di Criscienzo et al. 
(2006), deals with a further class of radial pulsators
observed
in similar metal poor stellar fields: Population II Cepheids. In
particular, we present the new non-linear and convective
pulsational models for short period variables, generally named BL Her stars. 

We have
investigated the instability strip and shown  
the dependence of the pulsation region on mass and luminosity, as well as  
the effect of varying the 
the mixing lenght
parameter.  We have presented a detailed atlas of
light/radial-velocity curves.  All these curves show a large variety of
shapes, which is perhaps the most striking feature of BL Her stars as
already suggested by  Moskalik \& Buchler (1994) on the basis
of non-linear but radiative models.

Theoretical light curves have been transformed in the UBVRIJK photometric
bands, using
the bolometric corrections and color-temperature relations provided by
Castelli, Gratton \& Kurucz (1997a,b). On this basis, mean magnitudes and colors
have been evaluated and
 some relations both for the boundaries of the instability strip and
for the dependence of the absolute magnitude on period, mass and color are
derived.

Finally, we have  calculated the period-radius relation and found that it is in
excellent agreement with the empirical relation by Burki \& Meylan and
consistent with the one holding for shorter periods for
RR Lyrae stars.

%________________________________________________________________
\begin{acknowledgements}
It is a pleasure to thank F. Caputo for useful discussions and for 
her critical reading of the early draft of this paper. Finalncial support for this study was provided by MIUR, under the scientif project ''On the evolution of stellar systems: a fundamental step toword the scientific exploitation of ``VST'' (P.I Massimo Capaccioli).
\end{acknowledgements}
%________________________________________________________________
%______________________________________________________________

\end{document}